# Transport through Andreev Bound States in a Graphene Quantum Dot

Travis Dirks, Taylor L. Hughes, Siddhartha Lal, Bruno Uchoa,
Yung-Fu Chen, Cesar Chialvo, Paul M. Goldbart, Nadya Mason

Department of Physics and Frederick Seitz Materials Research Laboratory
University of Illinois at Urbana-Champaign, Urbana, IL 61801, USA

**Andreev reflection—where an electron in a normal metal backscatters off a superconductor into a hole—forms the basis of low energy transport through superconducting junctions. Andreev reflection in confined regions gives rise to discrete Andreev bound states (ABS), which can carry a supercurrent and have recently been proposed as the basis of qubits.[1-3] Although signatures of Andreev reflection and bound states in conductance have been widely reported[4], it has been difficult to directly probe individual ABS. Here, we report transport measurements of sharp, gate-tunable ABS formed in a superconductor-quantum dot (QD)-normal system, which incorporates graphene. The QD exists in the graphene under the superconducting contact, due to a work-function mismatch.[5, 6] The ABS form when the discrete QD levels are proximity coupled to the superconducting contact. Due to the low density of states of graphene and the sensitivity of the QD levels to an applied gate voltage, the ABS spectra are narrow, can be tuned to zero energy via gate voltage, and show a striking pattern in transport measurements.**

Most previous work on superconductor (SC)-graphene structures has focused on the nature of the supercurrent in well-coupled Josephson junctions.[7-10] Superconductor-quantum dot hybrids in graphene have not been studied, although recent work has predicted[11-14] and demonstrated[15] that ABS can be isolated by coupling them to discrete QD energy levels. However, the ABS peaks in previous SC-QD experiments were strongly broadened, either by the large lead density of states[15] or by the lack of a tunnel barrier.[16] In the work described in this Letter, sharp subgap conductance peaks are obtained by tunneling into a proximity coupled QD formed within graphene, a high-mobility zero-gap semiconductor.[17] We focus on the two lowest energy conductance peaks that occur below the superconducting gap, and show that they are a signature of transport via ABS. The spectral pattern of these peaks as a function of gate and bias voltage is consistent with a simple theoretical model of ABS spectra presented below, and can be accurately fit with a more detailed microscopic calculation.

The data shown in this Letter were taken from one single-layer graphene device (Sample A) and one multilayer device (Sample B, approximately 10 layers thick). Similar behavior was seen in three other devices (two single-layer and one bilayer). As the features are robust upon adding layers, it is evident that a precise Dirac-point band-structure is not a requirement. The sample geometry and measurement circuit are shown in Fig. 1a; the location of the quantum dot which forms beneath the SC probe is depicted in Fig. 2a (discussed below). The devices consist of normal-metal end contacts (Cr/Au) and SC tunnel probes[18, 19] (Pb/In); the tunneling resistances through the SC probes were typically 10-100 times larger than the end-to-end resistances.

In all samples, the charge neutral point, as seen in the end-to-end conductance vs. backgate measurements, shows a large offset to the positive gate side (see Figs. 1c and 1d for samples A and B, respectively). For example, Sample A shows an asymmetric cone around the Dirac point at $V_g \sim +17.5$ V. Both effects have been predicted[5] and observed,[6] and are consistent with a work function mismatch ($\Delta W$) at the metal-graphene interface, which results in a transfer of charge that equalizes the surface potentials.[5] Because of its low density of states, graphene is efficiently doped by this charge transfer. The Cr/Au-graphene interface of the end contacts is dominated by the work function of the Au[20] so ΔW ranges from 0.14 to 1.04 eV (see Methods for calculation). The positive ΔW indicates hole (i.e., p-type) doping of the graphene under the end leads.

The superconducting (SC) tunnel probe also leads to a local doping of graphene. However, for the Pb-graphene interface, ΔW is -1.15 to -0.25 eV and is negative, implying local electron (n-type) doping. As illustrated in Figs. 1b and 2a, the ΔW doping generates a potential well underneath the SC tunnel probe, which acts as a confining potential for a QD – i.e., a pn junction. Quantum dots formed by pn junctions have been observed in carbon nanotubes,[21] via work function doping, and in graphene,[22] where Klein tunneling through smooth barriers can lead to the confinement of carriers.[23, 24] The QD is also proximity coupled to the superconducting lead. The interplay of the resulting Andreev reflections with Coulomb charging effects gives rise to low-energy ABS (see Fig. 2b). The ABS appear as subgap conductance peaks in our tunneling conductance measurements. Parallel experimental results demonstrating individual Andreev bound states within a carbon nanotube quantum dot have recently been obtained.[25]

In Fig. 3a we show typical tunneling conductance measurements at a fixed gate voltage. The conductance is dominated by the characteristic BCS shape of the superconducting DOS of Pb, with the expected gap 2Δ = 2.6 meV. Small peaks occur inside the gap, where one might expect conductance to be suppressed exponentially with the tunnel barrier thickness.[26] No conductance is observed around zero-bias, implying that the tunnel barrier is not leaky. The two lowest-energy sub-gap peaks are symmetric in bias voltage and have strong temperature dependence; the peaks decrease in amplitude as the temperature is raised to ~ 0.8K, above which the amplitude remains constant. This behavior is consistent with the temperature dependence of Coulomb blockade peaks in a crossover from a quantum to a classical dot regime.[27] In addition to the lowest-energy subgap peaks, we observe an oscillatory contribution above and below the gap. The oscillations are due to geometric resonances between the end contacts and ABS states in the QD (discussed further in SOM); if the QD could be connected directly to the Cr/Au leads, the lowest-energy subgap peaks would still appear, but not the oscillations.[25] The oscillations can be clearly distinguished from the lowest energy sub-gap peaks by very different gate-voltage dependence, as discussed below.

Figure 4a shows a 2D map of conductance vs. bias and gate voltage for sample A. The lowest energy subgap peaks display a striking gate- and bias-voltage dependent pattern. Near zero gate voltage, the conduction peaks start to emerge from the SC gap edge. As the gate voltage becomes more negative, the peaks move toward zero bias and cross at $V_g \approx -7V$. As the gate voltage is further decreased, the peaks split and then begin to converge again below $V_g \approx -10V$. This pattern can be qualitatively explained as resonant transport through ABS levels (see Fig. 2b for a schematic); the levels can be calculated from a simple phenomenological model (see below) and quantitatively fit with detailed transport calculations (see SOM), as shown in Fig. 4b. The correspondence between the calculation and the data for the lowest-energy sub gap peaks is remarkable.

The appearance of subgap conductance peaks requires a competition between the charging energy ($U$) and the effective superconducting pairing ($\Delta_{\text{eff}}$) acting on the QD. We distinguish between two physical regimes: (i) $U \ll \Delta_{\text{eff}}$ and (ii) $U \gg \Delta_{\text{eff}}$. In regime (i), the spin-up and spin-down states of the QD are nearly degenerate. As these levels are gated to within $\Delta_{\text{eff}}$ of the Fermi energy of the SC, they are occupied by paired electrons/holes, and the QD effectively becomes incorporated as part of the SC interface.[27] The conductance is then BTK-like and thus suppressed inside the gap, as in SC-normal interfaces having large tunnel barriers.[26] In contrast, for regime (ii), the charging energy dominates, and the spin-up and down states are widely split in energy, promoting pair-breaking. The QD then acts like a normal metal, and ABS are formed from the discrete QD states, due to Andreev reflections at the SC-QD interface. Resonant transport through the ABS levels leads to the observed subgap conductance peaks (see Fig. 2a). The clear observation of the subgap features in the data suggests that our measurements are taken in regime (ii). Additional evidence for the existence of ABS in our system is provided via a calculation based on a microscopic Hamiltonian that describes a graphene quantum well that is proximity coupled to a superconducting lead (see SOM).

A phenomenological model that considers the effect of the SC proximity coupling on a single pair of spin-split QD states explains the lowest energy ABS physics. The effective Hamiltonian for a proximity coupled QD is:

$$H = (\varepsilon_\uparrow - E_{\text{shift}})c_\uparrow^\dagger c_\uparrow + (\varepsilon_\uparrow + U - E_{\text{shift}})c_\downarrow^\dagger c_\downarrow + \Delta_{\text{eff}} c_\downarrow^\dagger c_\uparrow^\dagger + \Delta_{\text{eff}}^* c_\uparrow c_\downarrow$$

where $E_{\text{shift}}$ represents the shift of the QD energy levels by the gate voltage and $\varepsilon_\uparrow$, $\varepsilon_\uparrow + U$ are the energies of the spin-split levels. For a QD of diameter $R \sim 100\ nm$ (i.e., roughly the size of the SC tunnel probe), the charging energy $U \sim e^2/\kappa R \sim 5$ meV (where e is the electron charge and $\kappa = \frac{1+\epsilon}{2}$ where $\epsilon \approx 4$ is the dielectric constant of $SiO_2$). The resulting ABS energy levels lie at $E_\pm = \frac{1}{2}\left(\pm U + \sqrt{4\Delta_{\text{eff}}^2 + (2\varepsilon_\uparrow - 2E_{\text{shift}} + U)^2}\right)$. For $U = 0$, the energies $E_\pm$ are larger than $|\Delta_{\text{eff}}|$ for all gate voltages, implying that the appearance of zero-bias subgap conductance features requires a nonzero charging energy. With increasing charging energy, a pair of zero-energy ABS appears at a critical value of $U$, $U_c = 2|\Delta_{\text{eff}}|$, upon tuning the gate voltage such that $E_{\text{shift}}(V_g) = \varepsilon_\uparrow + \Delta_{\text{eff}}$. At this gate voltage and $U_c$, the bound states are equal-amplitude superpositions of particle and hole states, and thus effectively charge-neutral. At zero temperature, a quantum critical point separates the superconducting ($U < U_c$) and Coulomb charging ($U > U_c$) regimes of the QD.

For $U > U_c$, ABS appear within the gap, with the gate-voltage dependence illustrated in Fig. 5. Comparison with Fig. 4a shows that we observe the low-energy bound states and that $E_-$ can be extracted from the data. For example, at $V_g$ = -10V, $E_-$~ 0.5meV. $U$ can also be extracted if both zero-bias crossing points are observed, though the accuracy is only as good as the knowledge of the gate capacitance (see SOM). From the fit in Fig. 4b, we extract $U$ ~ 7.7 meV. Interestingly, Fig. 5 also shows that there are two gate voltages at which zero-energy bound states exist; the second crossing point lies just outside the gate-voltage range of our measurements. We note that for the transport calculations, sharp ABS resonances could only be found when the normal(superconducting) lead coupled to the QD had relatively low(high) DOS; this implies that a low-density "lead" such as graphene is indispensible for obtaining narrow ABS spectra.

In conclusion, we have reported direct observation of sharp, gate-tunable Andreev bound state spectra through transport experiments. The ABS peaks are resolved through a combination of a QD confined via a pn-junction in graphene, the low density of states in graphene, and the large tunneling barrier that insulates the QD from the SC lead. We have used a simple phenomenological model to explain the emergence of the ABS modes in terms of a competition between the superconducting-pairing and charging energies in the QD. This insight suggests the possibility of tuning through a quantum phase transition, which separates the two regimes, e.g., by changing the size of the QD. Finally, the sharp resolution of the levels—which have decoherence times inversely proportional to their width—may open a route towards applications, such as the design of quantum computing qubits in graphene-SC heterojunctions[2, 3].

This research was supported by the DOE-DMS under grant DE-FG02-07ER46453 through the Frederick Seitz Materials Research Laboratory at the University of Illinois at Urbana-Champaign, and partly carried out in the MRL Central Facilities (partially supported by the DOE under DE-FG02-07ER46453 and DE-FG02-07ER46471). TLH acknowledges the NSF under grant DMR-0758462 and the Institute for Condensed Matter Theory at UIUC, and BU acknowledges the DOE under grant DE-FG02-91ER45439. We acknowledge useful conversations with J. Maciejko.

Figures:

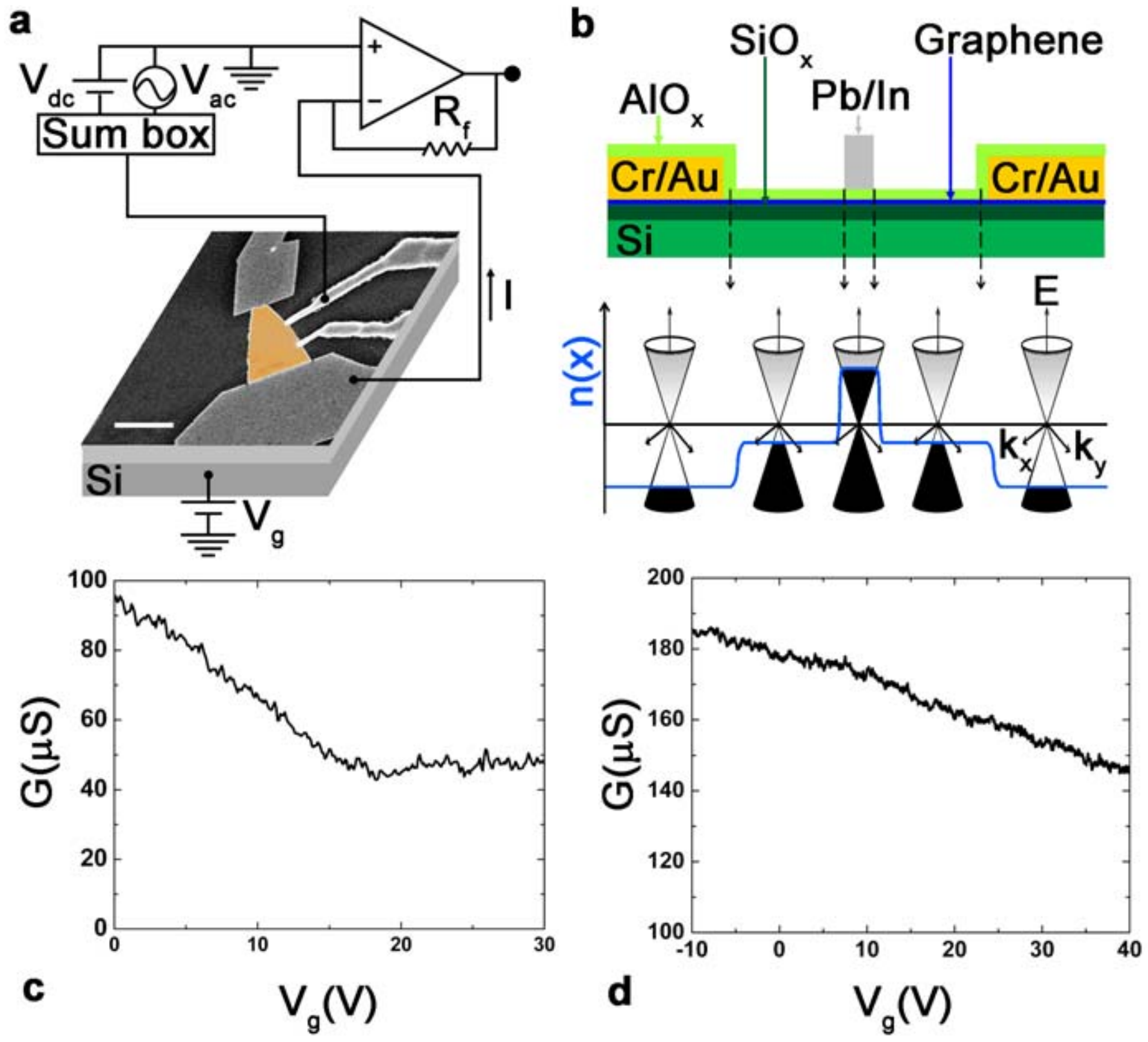

Figure 1. a) SEM micrograph of a device with overlaid measurement circuit. Graphene is false colored orange, large end contacts are Cr/Au, and middle tunnel probes are Pb/In. Scale bar is 5 μm. b) Above, side-view schematic of device. Below, illustration of doping profile as a function of position along the device (blue line) and Dirac cones showing location of Fermi level. Square well under tunnel probe shows where p-n junctions create confining potential for quantum dot. c) End-to-end conductance vs. backgate for sample A (single layer), displaying the Dirac cone. d) End-to-end conductance vs. backgate for sample B (multilayer). The asymmetry in (c) and (d) shows that the bulk graphene is p- doped by the backgate.

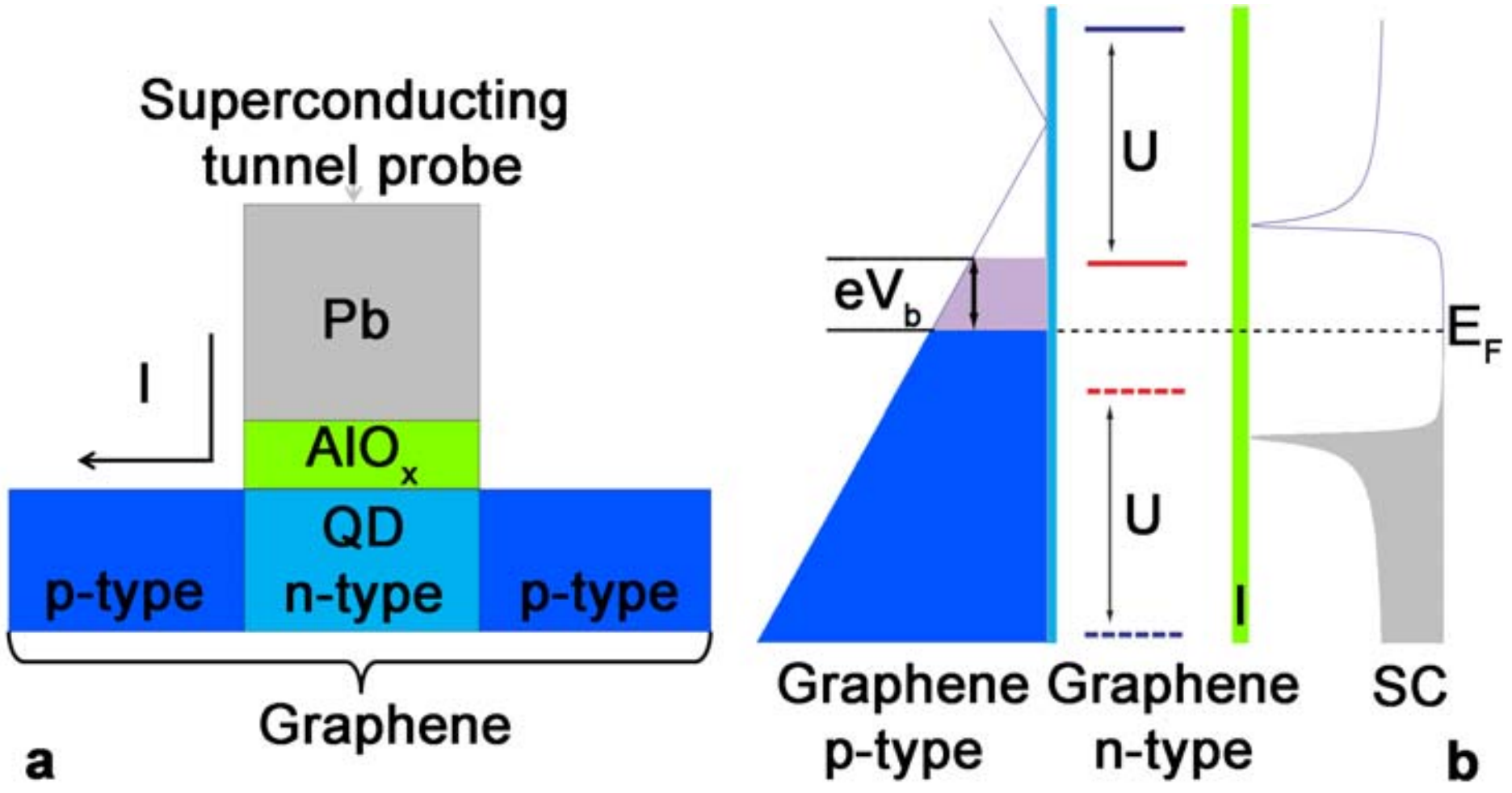

Figure 2: a) Schematic of the quantum dot formed in graphene by a work function mismatch at the Pb interface. b) Energy level schematic of the graphene-QD-superconductor system. The DOS of the p-type graphene and SC tunnel probe is shown on the left and right, respectively, with filled states indicated. $AlO_x$ tunnel barrier is indicated in green on the right and the p-n junction is indicated in light blue on the left. Blue/red energy levels refer to Andreev bound states. The solid(dashed) lines represent states which have dominant particle(hole) character. The bias voltage, $V_b$, is shown tuned to allow resonant subgap conduction.

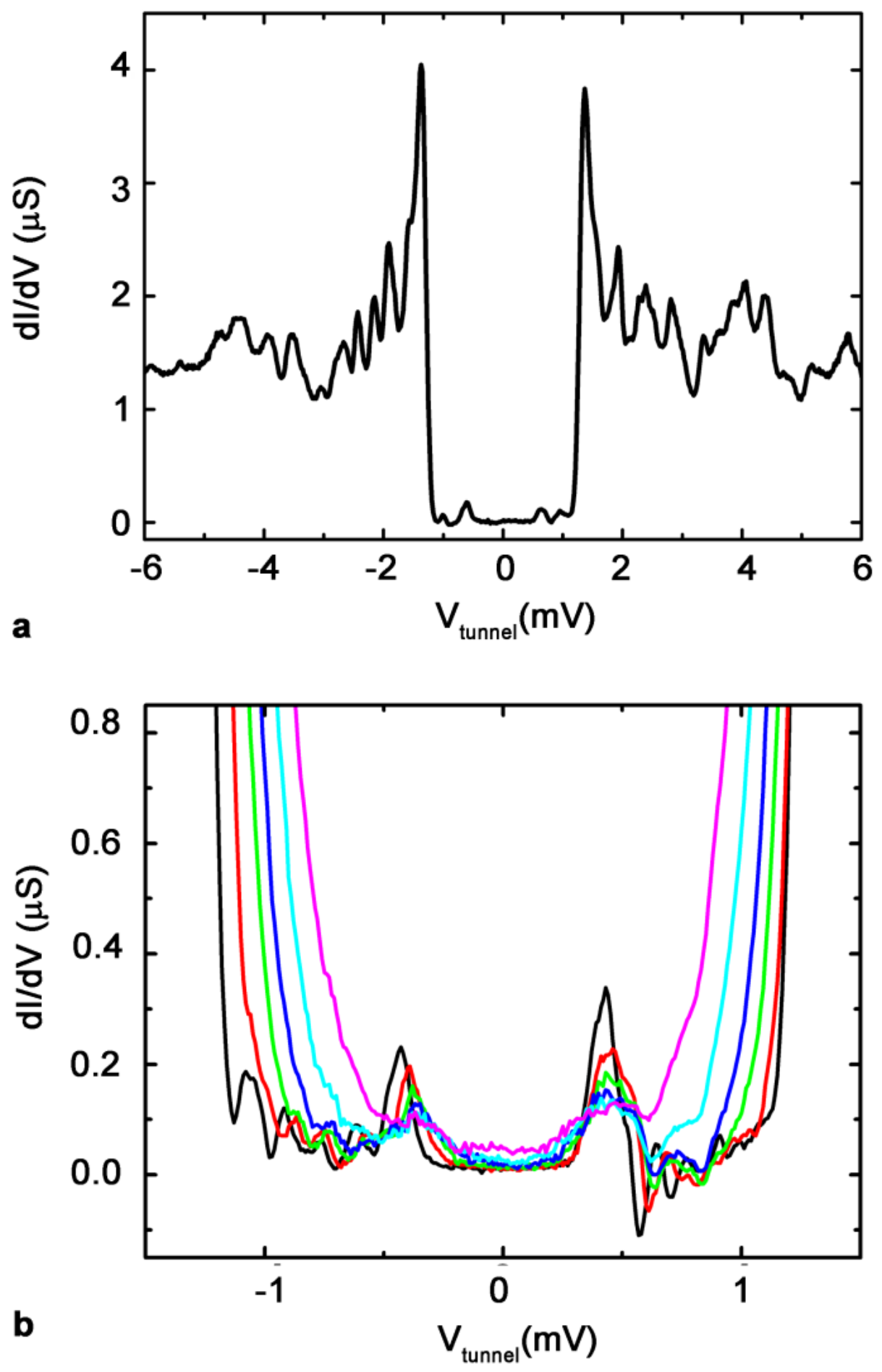

Figure 3: a) Tunneling differential conductance vs. bias voltage (setup as in Fig. 1a), for the multilayer graphene (sample B). Large conductance oscillations outside the gap are likely Fabry-Perot interference effects. Similar oscillations and sub-gap peaks are seen in sample A. Two ABS peaks are visible inside the SC gap. b) Temperature dependence of the subgap peaks. The temperature from the widest SC gap is 0.26 K, 0.45 K, 0.67 K, 0.86 K, 1.25 K, and 1.54 K. The peaks decrease in amplitude and increase in breadth as temperature is increased to ~ 0.8K, then remain constant; this is consistent with a cross-over from a quantum to a classical dot regime.

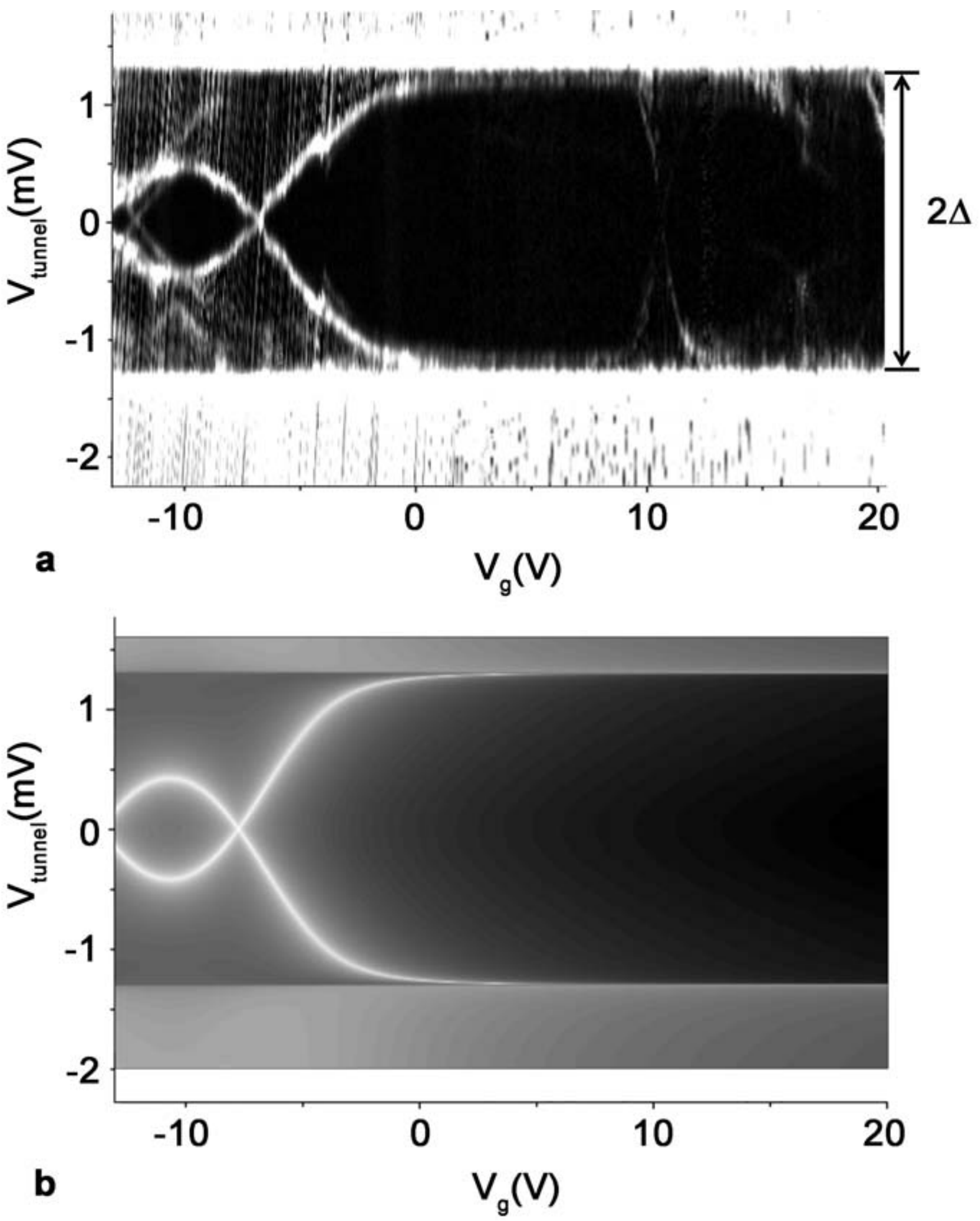

Figure 4: a) 2D map of tunneling differential conductance vs. backgate voltage (x-axis) and bias voltage (y-axis) on a log scale for the single-layer device (sample A). Bright white lines inside the gap (marked as 2Δ) are subgap peaks, or ABS, which are symmetric about zero bias and gate-dependent. b) A fit of the conductance data from the detailed transport calculations for a quantum dot with two levels, a finite charging energy, and with couplings to normal metal and superconducting leads (see main text and SOM).

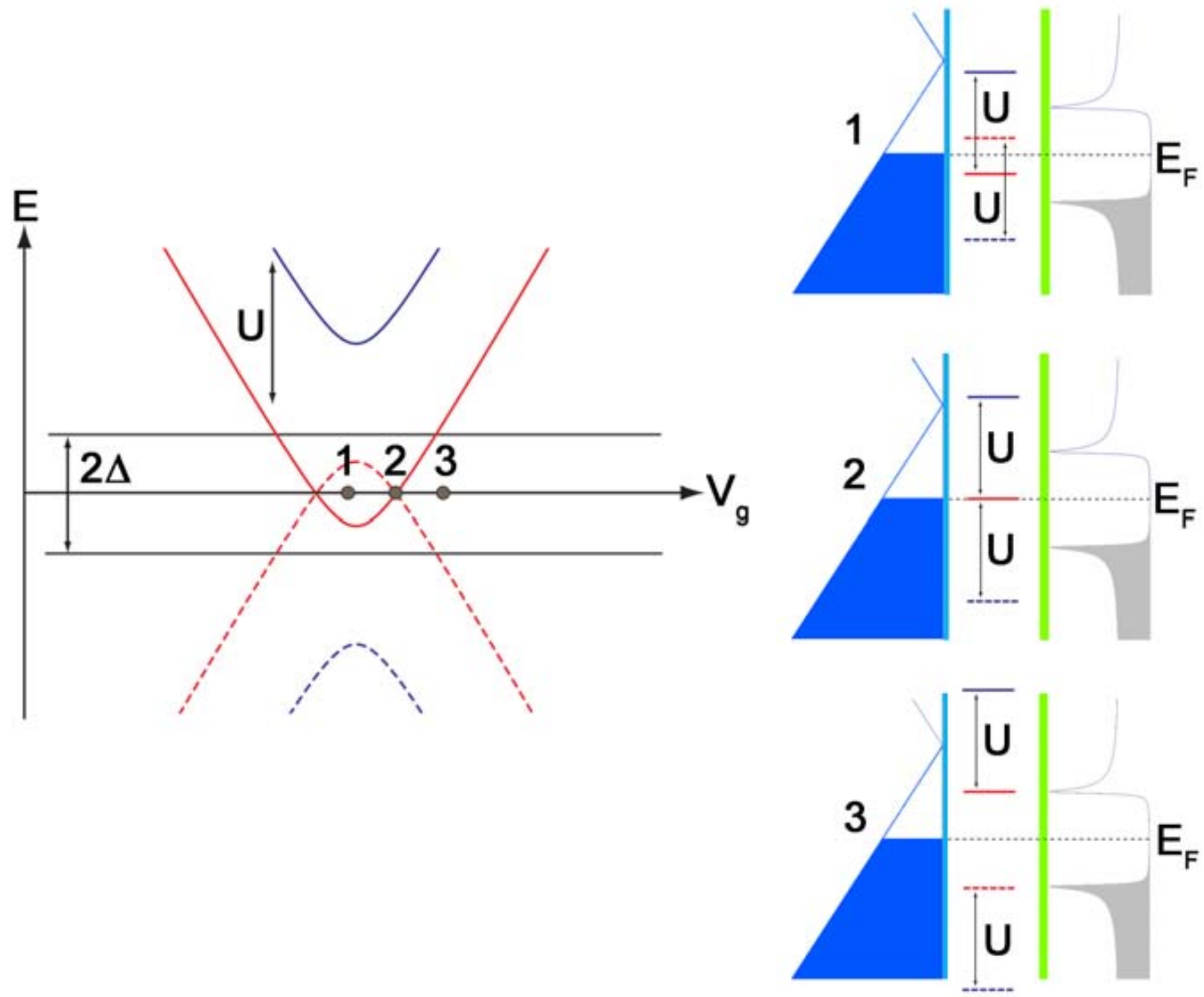

Figure 5: On left: Energy diagram showing the evolution of Andreev bound state levels in a quantum dot with varying gate voltage. $U$ is the charging energy of the quantum dot and $\Delta$ is the superconducting gap of the tunnel probe. On right: Three resonant tunneling diagrams corresponding to the three different gate voltages marked on the left-side diagram. The solid(dashed) lines represent states which have dominant particle(hole) character. The bias voltages required for conductance through the ABS are the energy differences between the bound state levels and $E_F$ of the superconductor. For point 1, one ABS (red) is below the SC gap edge and one (blue) is above; this gives two subgap peaks (red levels) in the conductance at finite (positive and negative) bias voltage. At point 2, an ABS is tuned to zero energy, which leads to a zero-bias conductance peak. At point 3, the ABS level is at the gap edge, which leads to subgap peaks which are pushed towards the gap edge.

**Methods**

The graphene samples were mechanically exfoliated onto highly doped Si substrates capped with 300 nm $SiO_2$. The graphene thickness was determined via optical microscopy, Raman spectroscopy, and atomic force microscopy. The devices consist of four electrodes on a piece of graphene (see Figure 1a). The two large end leads are Cr/Au and the narrower middle probes are Pb/In. The end electrodes were patterned by conventional electron beam lithography and electron beam evaporation of 2 nm Cr and 50 nm Au. The chips were then annealed in $H_2$ and Ar at 300C for 2 hours. Next, the devices were covered in 12 layers $AlO_x$ via Atomic Layer Deposition, and finally, 200 nm wide SC tunnel probes were patterned by conventional electron beam lithography and thermal evaporation of 200 nm Pb and 30 nm In. The tunneling resistances through the SC probes, $R_{\text{tunnel}} \sim 200 – 500$ KΩ, are typically 10-100 times larger than the graphene's end-to-end resistances, $R_{\text{end-to-end}} \sim 5 – 20$ KΩ. Sample A is a single-layer graphene device with a distance between the two end contacts $L \sim 4.2$ µm, a width $W \sim 1.5$ µm, and a SC probe junction size of ~ 0.2 x 0.2 µm (0.3 x 0.2 µm). Sample B is a multi-layer device having $L$~6.4 µm, $W \sim 0.8$ µm, and a SC probe junction ~ 0.3 x 0.2 µm. Measurements were performed in a Helium-3 cryostat using standard ac lock-in techniques. The measurement set-up is shown in Figure 1a. Tunneling differential conductance measurements were performed by applying a sum of dc bias voltage $V_{dc}$ and ac excitation voltage $V_{ac}$ to the superconducting tunnel probe, and a voltage $V_g$ to the back gate, while measuring the differential conductance $dI_{ac}/dV_{ac}$ at one of the graphene end contacts as illustrated in Figure 1a.

The sign of the charge transfer due to work-function-mismatch doping is the same as that of the mismatch $\Delta W = W_m – W_g - W_c$ , where $W_m$ is the work function of the metal, $W_g = 4.5$ eV is the work function of the graphene,[5] and $W_c \sim$ 0-0.9 eV is an effective potential that arises from metal-graphene chemical interactions.[5] For the Cr/Au-graphene interface of the end contacts, $W_{Au}$ = 5.54 eV so ΔW = 0.14 – 1.04 eV. For the SC tunnel probe-graphene interface, $W_{Pb}$ = 4.25 eV so ΔW = -1.15 – -0.25 eV. For the work function of the metals, we use standard theoretical values (see CRC handbook, 2008), which are consistent with known experimental results.

# Supporting Online Material:Transport through Andreev Bound States in a Graphene Quantum Dot

Travis Dirks, Taylor L. Hughes, Siddhartha Lal, Bruno Uchoa,
Yung-Fu Chen, Cesar Chialvo, Paul M. Goldbart, Nadya Mason
*Department of Physics and Seitz Materials Research Laboratory,*
*University of Illinois, 1110 West Green Street, Urbana 61801, USA.*

## I. TRANSPORT CALCULATIONS

The model Hamiltonian we used to fit the data in the manuscript is a quantum dot coupled to one normal metal lead on the left and one $s$-wave superconducting lead on the right. We construct the model Hamiltonian in the Bogoliubov-de Gennes formalism [1]. The total Hamiltonian is given by

$$H = H_L + H_{L-QD} + H_{QD} + H_{QD-R} + H_R \tag{1}$$

where

$$H_L = -t_L \sum_{\langle i,j\rangle,\sigma} c^\dagger_{Li\sigma} c_{Lj\sigma} \tag{2}$$

is the Hamiltonian for a semi-infinite metallic lead, with $\langle ij\rangle$ summed over nearest neighbor lattice sites on a three dimensional square lattice, and $\sigma =\uparrow,\downarrow$ summed over the spins. On the superconductor (SC) side,

$$H_R = -t_R \sum_{\langle i,j\rangle,\sigma} c^\dagger_{Ri\sigma} c_{Rj\sigma} + \sum_i \left(\Delta c^\dagger_{Ri\uparrow} c^\dagger_{Ri\downarrow} + \text{h.c.}\right), \tag{3}$$

is the Hamiltonian for a semi-infinite SC lead, where $\Delta$ is the SC gap, and

$$H_{QD} = [\epsilon_\uparrow - \mu(V_g)]\, d^\dagger_\uparrow d_\uparrow + [\epsilon_\uparrow + U - \mu(V_g)]\, d^\dagger_\downarrow d_\downarrow, \tag{4}$$

describes the Hamiltonian for a two-state quantum dot with charging energy $U$ and chemical potential $\mu(V_g)$ as a function of gate voltage. The connection between the left and right leads to the QD is given by the two terms:

$$H_{L-QD} = -t_{L-QD} \sum_\sigma \left(c^\dagger_{LN\sigma} d_\sigma + d^\dagger_\sigma c_{LN\sigma}\right) \tag{5}$$

$$H_{QD-R} = -t_{QD-R} \sum_\sigma \left(c^\dagger_{R1\sigma} d_\sigma + d^\dagger_\sigma c_{R1\sigma}\right). \tag{6}$$

The expression of the current can be found for instance in Ref.[2]. The total current is $I = I_\uparrow + I_\downarrow$, where

$$I_\uparrow = \frac{e}{h}\int d\omega \mathrm{Tr}\left[\Gamma_{L\uparrow}\left[G^r\Gamma_R G^a\right]_{11}(f_\uparrow - f) + \Gamma_{L\uparrow}G^r_{12}\Gamma_{L\downarrow}G^a_{21}(f_\uparrow - f_\downarrow)\right] \tag{7}$$

$$I_\downarrow = -\frac{e}{h}\int d\omega \mathrm{Tr}\left[\Gamma_{L\downarrow}\left[G^r\Gamma_R G^a\right]_{22}(f_\downarrow - f) + \Gamma_{L\downarrow}G^r_{21}\Gamma_{L\uparrow}G^a_{12}(f_\downarrow - f_\uparrow)\right], \tag{8}$$

with $f(\omega) = \left[1 + e^{(\hbar\omega-\mu)/k_BT}\right]^{-1}$ the Fermi distribution, $T$ the temperature, and $f_\uparrow(\omega) = f(\omega - eV)$, $f_\downarrow(\omega) = f(\omega + eV)$ describe the excited state distributions. $G^{r/a}(\omega)$ is the retarded/ advanced Green's functions of the quantum dot

$$G^{r/a}(\omega) = \frac{1}{\omega - H^{BdG}_{QD} - \Sigma^{r/a} \pm i0^+} \tag{9}$$

where

$$H^{BdG}_{QD} = \frac{1}{2}\begin{pmatrix} H_{QD} & 0 \\ 0 & -H^T_{QD} \end{pmatrix}. \tag{10}$$

is the BdG matrix Hamiltonian of the QD, and $\Sigma^{r/a}$ is the retarded/advanced self-energy due to the coupling to the leads.

The indexing, as in, for example, $[G^r\Gamma_R G^a]_{11}$ indicates the matrix block in Nambu-space *i.e.* the 11 index means the particle-particle block. The self-energies due to the semi-infinite leads are calculated self-consistently using the method of Ref. [3]. The broadening functions are related to the self-energies via

$$\Gamma_{L\uparrow}(\omega) = i\left(\Sigma^r_L - \Sigma^a_L\right)_{11} \tag{11}$$

$$\Gamma_{L\downarrow}(\omega) = i\left(\Sigma^r_L - \Sigma^a_L\right)_{22} \tag{12}$$

and similarly for $\Gamma_R$. For more details on the notation see Ref. [2].

In the low-temperature, small bias voltage limit we can approximate the conductance by

$$G_{LR}(\omega) = -\frac{e^2}{h}\left\{\mathrm{Tr}\left[\Gamma_{L\uparrow}\left[G^r\Gamma_R G^a\right]_{11} + 2\Gamma_{L\uparrow}G^r_{12}\Gamma_{L\downarrow}G^a_{21} + \Gamma_{L\downarrow}\left[G^r\Gamma_R G^a\right]_{22} + 2\Gamma_{L\downarrow}G^r_{21}\Gamma_{L\uparrow}G^a_{12}\right]\right\} \tag{13}$$

which can be compared with the conductance maps in the experimental data. For the fitting procedure we chose the following parameters: $\Delta = 1.3$ meV, $U = 7.7$ meV, $\epsilon_\uparrow = -32.4$ meV, $t_L = t_R = 1.0$ eV, $t_{L-QD} = 0.8$ eV, $t_{QD-R} = 0.0495$ eV and

$$\mu(V_g) = A(V_g - 17.5). \tag{14}$$

The values of $U$ and $\epsilon_\uparrow$ depend very sensitively on the capacitance scale factor relating the gate voltage to the energy shift of the quantum dot levels. Our estimate for this quantity $A = \frac{1.01\,\mathrm{meV}}{V}$ is based on an estimate for the gate capacitance for graphene on a 300 nm $SiO_2$ substrate. However, the QD region is in very close proximity to the tunnel probe (separated by only a 1 nm barrier) so it is likely that the gate is further screened and the capacitance scale factor would be reduced possibly even as much as an order of magnitude. Given this wide range, the values of $U$ and $\epsilon_\uparrow$ should only be taken as rough estimates for the actual values because they are only as accurate as the knowledge of the effective capacitance. The value for $\Delta$ used is the pairing potential in the lead (Pb) not the quantity $\Delta_{eff}$ discussed in the manuscript. $\Delta_{eff}$ is primarily a function of $\Delta, t_{QD-R}$ and energy. Additionally note that no physical insight can be gained from the value of the parameter $t_{L-QD}$. This parameter has no effect on the shape of the conductance curves and only affects the heights of the conductance peaks. The value we used was chosen for aesthetics, *i.e.* to give a good contrast ratio in the conductance-map.

The above calculation describes the behavior of the two lowest-energy subgap peaks. However, our data also shows oscillations on top of the typical SIN conductance background. These oscillations can be clearly distinguished from the lowest energy sub-gap conductance peaks by their strikingly different gate and bias voltage dependencies, as can be seen in Fig. 4a of the main manuscript. In addition, as can be seen in the comparison between Figs. 4a and 4b, the lowest-energy ABS peaks can be analyzed and fit independently of the oscillations.

In our data, we have evidence that the oscillations arise from the coupling of the QD to resonances in the graphene cavity between the QD and the Cr/Au lead. The energy level spacing which is expected for resonances in a 3.5 $\mu$m long cavity in graphene is of the order of 0.5 meV; this is consistent with the characteristic spacing between oscillations observed experimentally. In addition, the oscillations displayed in the conductance plots in Fig. 4a show a pattern of very narrowly spaced straight lines, which can be heuristically described by the formula

$$AV_{\text{gate}} - E_{\text{bias}} = \pi n v / L \,, \tag{15}$$

for $n$ as integers, where $v = 6$ eVÅ is the Fermi velocity in graphene, $V_{\text{gate}}$ is the gate voltage, $E_{\text{bias}}$ is the bias, $A$ is the dimensionless scale-factor which converts between applied gate voltage and the shift of the chemical potential (note its appearance in Eq. 14 above) and $L$ is a confining length scale, of the order of the size of the cavity itself. The slope in the pattern of the oscillations for bias vs. gate voltage is $A \approx 0.8\ meV/V$, which is a number comparable to the capacitance of the QD used to fit the lowest energy ABS conductance peaks in the transport calculation. This argument indicates that the oscillations have a different origin than the lowest-energy subgap conductance features, and are likely caused by geometrical resonant paths in the cavity.

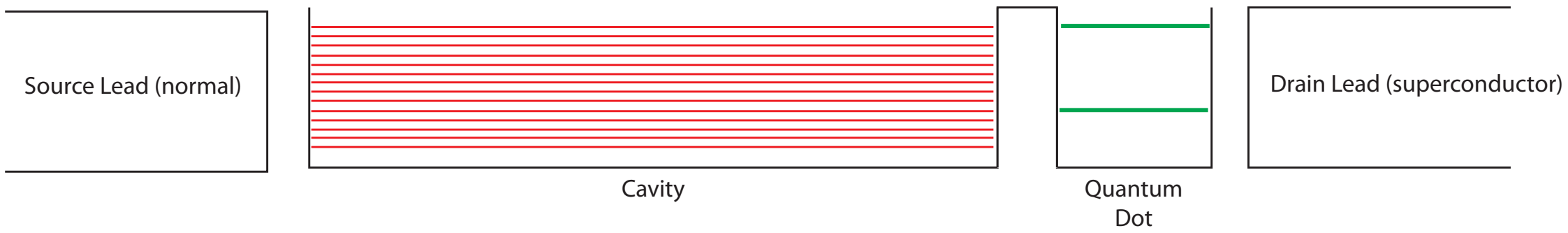

Figure 1: Schematic diagram of a quantum dot coupled to a cavity of resonances. We will consider transport through this system coupled to normal metal and superconducting leads.

Thus, in order to describe the conductance oscillations in the data, one needs in addition to model how the QD attaches to the normal lead. In the experiment, the QD is not attached directly to it, but is in reality connected to a mesoscopic graphene flake which is in turn attached to the normal metal lead. We can use a simple 1D model as a proof of concept which captures some of the basic features of the conductance oscillations due to geometric resonances. We will use a quantum dot coupled to a superconducting lead (on the right) and a cavity (on the left) which is in turn is coupled to a normal metal lead. The model geometry is shown in Fig. 1. The model Hamiltonian is

$$H = H_L + H_{L-C} + H_C + H_{C-R} + H_R \tag{16}$$

where $H_L, H_R$ are given in Eqs. 1,2 respectively

$$H_C = -t_C \sum_{\langle i,j \rangle,\sigma} c^\dagger_{Ci\sigma} c_{Cj\sigma} + \sum_i c^\dagger_{Ci\sigma} (U_i \sigma^z_{\sigma\sigma'} + V_i) c_{Ci\sigma'} \tag{17}$$

$$H_{L-C} = -t_{L-C} \sum_\sigma \left( c^\dagger_{LN\sigma} c_{C1\sigma} + c^\dagger_{C1\sigma} c_{LN\sigma} \right) \tag{18}$$

$$H_{C-R} = -t_{C-R} \sum_\sigma \left( c^\dagger_{R1\sigma} c_{CN\sigma} + c^\dagger_{CN\sigma} c_{R1\sigma} \right). \tag{19}$$

The position dependent charging energy $U_i$ and is only non-zero in the region of the small confined quantum dot. The electrostatic potential $V_i$ is chosen to confine the quantum dot with barriers but is zero everywhere else. Thus $H_C$ represents both the cavity and the quantum dot when $U_i$ and $V_i$ are set correctly.

The model consists of the same elements as the simple case above with an intermediate Hamiltonian serving as a cavity. We model the cavity Hamiltonian as a “quantum dot" without a charging energy. The discrete levels in the cavity are much more closely spaced than those in the quantum dot to mimic a large region. Additionally we The coupling between the cavity and quantum dot is weakened by the barrier between them which serves also to confine the quantum dot.

At a fixed gate voltage we see a typical conductance slice as a function of bias voltage in Fig. 2. These features resemble our data. It is very difficult to quantitatively, or even qualitatively model these types of oscillations without accurately characterizing the geometrical scattering pathways of each sample and the tunnel couplings to the leads.

## II. WKB CALCULATION OF ANDREEV BOUND STATES

Our calculation of the Andreev bound states (ABS) in a quantum well that is proximity coupled to a SC lead, as shown in Fig. 3, is based on the standard BdG Hamiltonian of a graphene quantum dot (QD), (see Ref. [1])

$$H_{BG} = \begin{pmatrix} H_g & \Delta \\ \Delta^* & -H_g \end{pmatrix},$$

where

$$H_g = -t \sum_{\sigma} \sum_{\langle ij \rangle} a^{\dagger}_{i,\sigma} b_{j,\sigma} + \text{h.c.} - \tilde{E}_F \sum_{\sigma} \hat{n}_{i,\sigma} + \hat{n}_{i,\uparrow} U \tag{20}$$

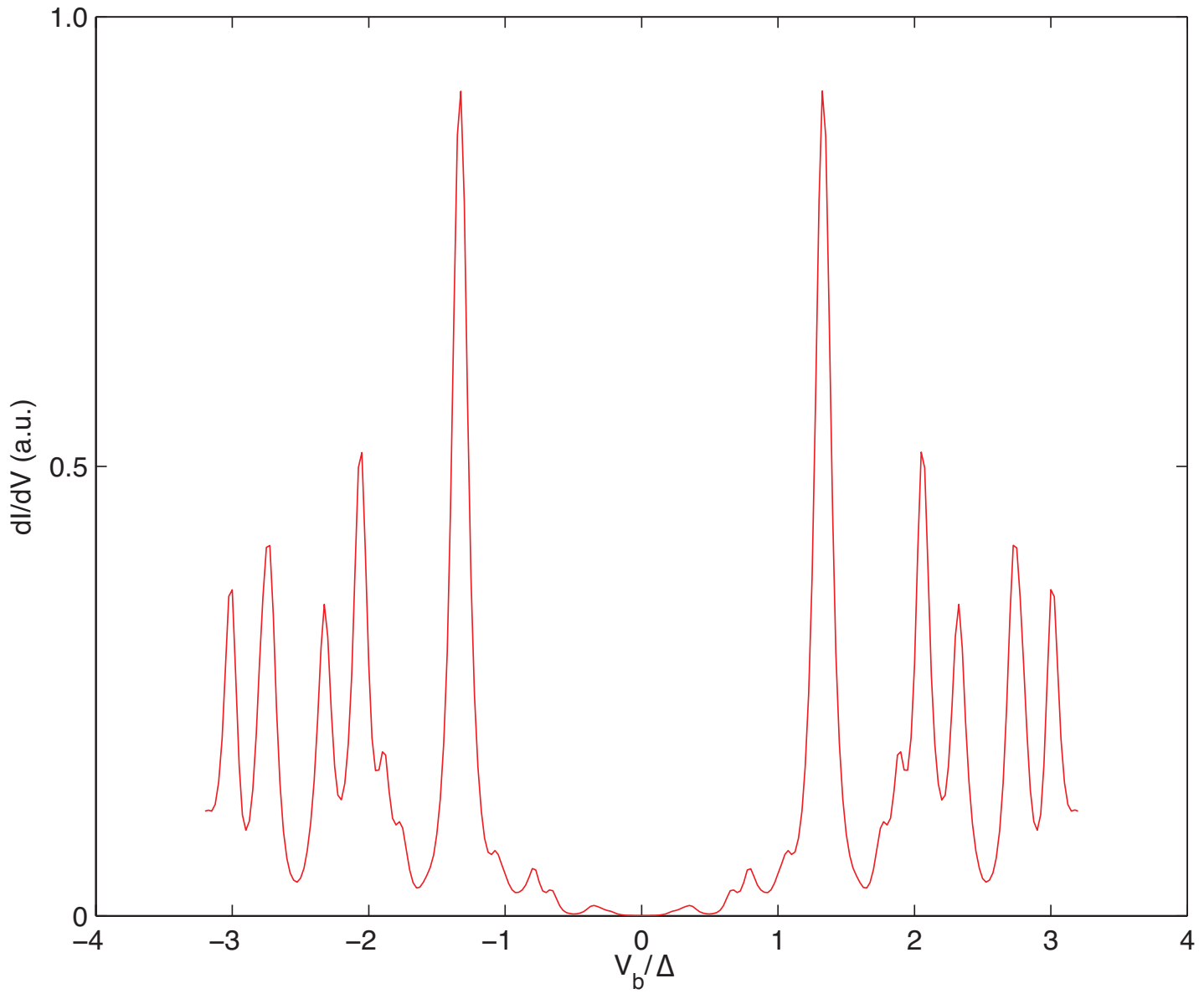

Figure 2: A conductance slice at fixed gate voltage as a function of bias voltage.

is the graphene tight-binding Hamiltonian in the presence of an effective charging energy $U$ that lifts the spin degeneracy, thus polarizing the spin in the QD, $a_{i,\sigma}$ and $b_{i,\sigma}$ are the fermionic operators on the graphene sublattice $A$ and $B$, $\hat{n}_i$ the particle density operator on site $i$, and $t \approx 3$ eV is the hopping amplitude. The parameter $\tilde{E}_F = E_F + CV_g + V_w(x)$ is an effective Fermi energy in the QD, where $E_F \approx -140$meV is the Fermi energy in graphene, $C \sim 10^{-3}$ is the effective capacitance of the back gate $V_g$, and

$$V_w(x) = -u_0 + \frac{4u_0}{L^2}\left(x + \frac{L}{2} + d\right)^2 \tag{21}$$

is the potential well of the QD, centered at $x_0 = -L/2 - d$, as shown in Fig.3. The energy $u_0$ gives the height of the well, $L \approx 200$ nm is its spatial extent, and $d \approx 1$ nm is the thickness of the insulating (I) barrier separating the QD from the SC.

We use this simple geometry to show that despite Klein tunneling, Andreev bound states (ABS) may form in a graphene quantum well proximity-coupled to a I-SC interface, as a result of the interplay of the superconductivity and the Coulomb charging energy in the dot. For Dirac particles, it is not the height but the slope of the barrier that results in the scattering and possible confinement of charge carriers. We note that as $k_F L$ is not a large number, the semiclassical calculation may not necessarily be a quantitatively accurate description, but it should, nevertheless, describe the qualitative features of ABS as a function of gate and bias voltages in a QD strongly coupled to a SC lead.

In the continuum limit, the Hamiltonian (20) becomes Dirac-like, i.e.

$$H_g = \sum_{\mathbf{k}} \left[\Psi^\dagger_{\mathbf{k},\downarrow}(-\hbar v \mathbf{k}\cdot\vec{\sigma} - E_F)\Psi_{\mathbf{k},\downarrow} + \Psi^\dagger_{\mathbf{k},\uparrow}(-\hbar v \mathbf{k}\cdot\vec{\sigma} - E_F + U)\Psi_{\mathbf{k},\uparrow}\right], \tag{22}$$

where $\Psi_{\mathbf{k},\sigma} \equiv (a_{\mathbf{k},\sigma}, b_{\mathbf{k},\sigma})$ is a two-component spinor, $\mathbf{k}\cdot\vec{\sigma} \equiv -i(\sigma_x\partial_x + \sigma_y\partial_y)$, and $\hbar v \approx 6$eVÅ is the Fermi velocity. Defining $\hat{H}_{g,\downarrow} = -\hbar v\mathbf{k}\cdot\vec{\sigma} - E_F$ and $\hat{H}_{g,\uparrow} = -\hbar v\mathbf{k}\cdot\vec{\sigma} - E_F + U$, the BdG equation in the spin symmetrized form becomes

$$\frac{1}{2}\begin{pmatrix} \hat{H}_{g,\uparrow} & 0 & 0 & \Delta \\ 0 & \hat{H}_{g,\downarrow} & -\Delta & 0 \\ 0 & -\Delta^* & -\hat{H}_{g,\uparrow} & 0 \\ \Delta^* & 0 & 0 & -\hat{H}_{g,\downarrow} \end{pmatrix}\begin{pmatrix} \Psi_\uparrow \\ \Psi_\downarrow \\ \Psi^\dagger_\uparrow \\ \Psi^\dagger_\downarrow \end{pmatrix} = E\begin{pmatrix} \Psi_\uparrow \\ \Psi_\downarrow \\ \Psi^\dagger_\uparrow \\ \Psi^\dagger_\downarrow \end{pmatrix}; \tag{23}$$

the eigenvalues are readily found to be

$$E_\pm = \alpha\frac{U}{2} \pm \sqrt{\left[vk + \eta\left(\frac{U}{2} - \tilde{E}_F\right)\right]^2 + |\Delta|^2}\,, \tag{24}$$

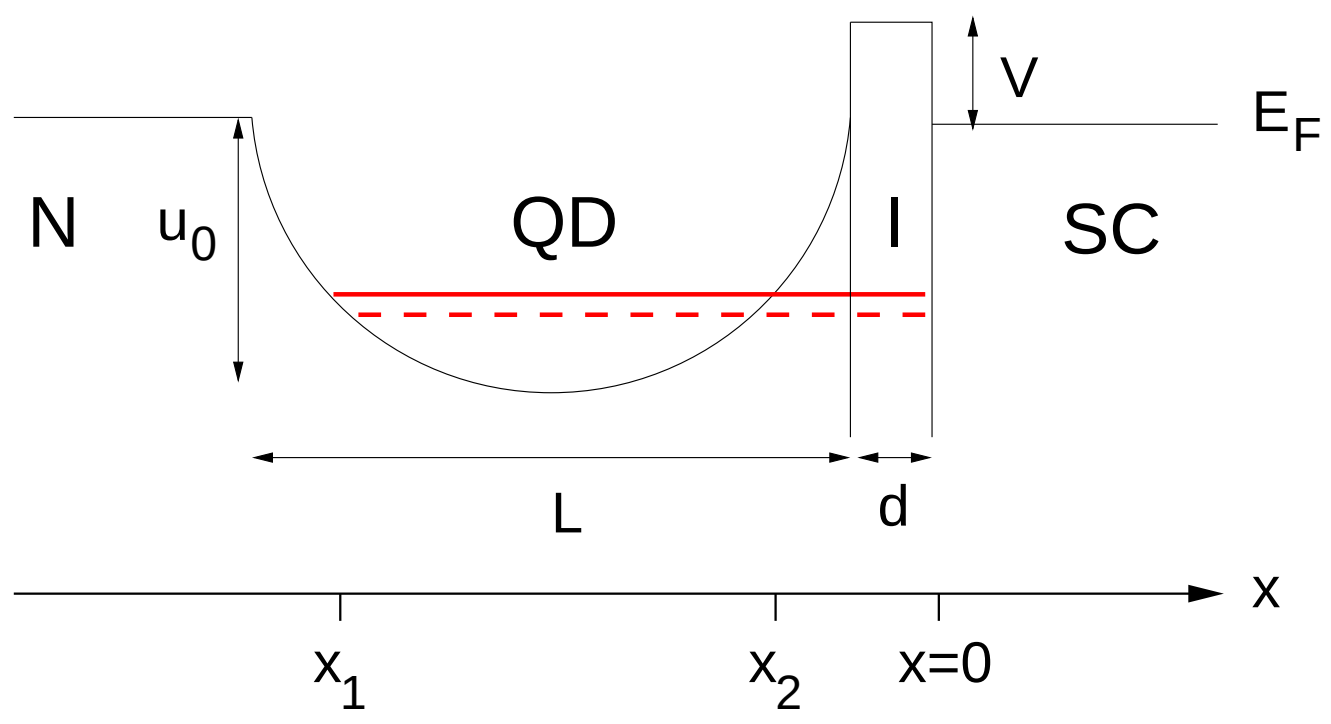

Figure 3: Schematic depiction of a normal(N)-QD-I-SC heterojunction in graphene. The Fermi level in the junction is pinned by the SC. The normal graphene region has a Fermi energy $E_F \approx -140$meV. The energy $u_0$ is the depth of the well, with $L \approx 200$nm in size. The solid and dashed red lines indicate the resonant paths made of admixtures of particles and holes that form the Andreev bound states.

where $\alpha = \pm$ labels the spin-split states, $\eta = +(-)$ labels the conduction (valence) bands of graphene, and $k \equiv \sqrt{k_x^2 + k_y^2}$.

When an electron is incident at the SC interface, it can propagate into the SC as a Cooper pair provided that a hole is Andreev reflected back into the well. As the particles bounce back and forth between the left hand side of the proximity coupled QD and the I-SC interface, they acquire a phase that can be described semiclassically by the line integral of the longitudinal momentum in a round-trip path, $\theta = \oint k_x(x)\mathrm{d}x$. The contribution of a particle traversing the well is set by the position of the two classical turning points, defined when $k_x = 0$,

$$x_{1,2} = x_0 \pm \frac{L}{2\sqrt{u_0}} \left[ \pm\eta\sqrt{\left(E - \alpha\frac{U}{2}\right)^2 - |\Delta|^2 - \eta v|k_y| - \tilde{U}} \right]^{1/2}, \tag{25}$$

where $\tilde{U} \equiv U/2 - E_F - CV_g - u_0$.

The total phase shift accumulated along one closed path (i.e., from $x_1$ to $x = 0$ and back), is $\theta = 2\varphi_{\mathrm{WKB}} + \varphi_{\mathrm{barrier}} + \varphi_{\mathrm{Andreev}}$, where $\varphi_{\mathrm{WKB}} = \int_{x_1}^{-d} k_x(x)\mathrm{d}x$, is the WKB phase across the well to the barrier, and[4]

$$\varphi_{\mathrm{barrier}} = [p_b(E) - p_b(-E)]\, d - \frac{1}{2}\,[\delta(E) - \delta(-E)] \tag{26}$$

is the phase-shift accrued in traversing the barrier of height $V$, where

$$p_b(E) = \frac{1}{\hbar v}\sqrt{(E + E_F - V)^2 - (\hbar v|k_y|)^2}\,, \tag{27}$$

and

$$\delta(E) = \arcsin\left(\frac{\hbar v|k_y|}{E + E_F - V}\right), \tag{28}$$

describe the momentum and the scattering phase-shifts for electrons [conversely, $p_b(-E)$and $\delta(-E)$ correspond to the case of holes]. The Andreev scattering phase-shift at the I-SC interface is [7]

$$\tan\varphi_{\text{Andreev}} = -\tan[\beta(E)]\,\frac{\cos\left\{[\delta(E)+\delta(-E)]/2\right\}}{\cos\left\{[\delta(E)-\delta(-E)]/2\right\}}\,, \tag{29}$$

where

$$\beta(E) = \begin{cases} \arccos\left(\frac{E}{|\Delta|}\right) & \text{for } E < |\Delta| \\ \\ -i\text{arccosh}\left(\frac{E}{|\Delta|}\right) & \text{for } E > |\Delta|\,. \end{cases} \tag{30}$$

The total phase-shift for a bound state satisfies the Bohr-Sommerfeld quantization conditions $\theta = \left(n+\frac{1}{2}\right)\pi\hbar$, for integer $n$. The lowest energy solution of this equation describes the Andreev bound state shown in Fig.4. Points to the left and right of the center of the inner diamond in that figure, at $V_g \approx 8.8$ V, correspond to bound-state solutions resulting from two distinct spin-split eigenstates of Eq. (24), respectively. Precisely at the edge of the diamond, the two solutions are degenerate, and the envelope resembles a Coulomb blockade diamond. Additional rounding of the diamond shape is expected from treating the SC order parameter in the QD self-consistently, as described in Ref.[5, 6]. Unlike those previous works, our work demonstrates from a microscopic perspective, the interplay between proximity-coupled superconductivity and Coulomb charging effects in a QD. The solutions shown in Fig.4 reveal the main subgap features observed in Fig.4a of the main manuscript, demonstrating the existence of Andreev bound states in a graphene quantum well proximity-coupled to an I-SC interface, when pair-breaking Coulomb charging effects compete with Cooper pairing.

Finally note that in the experimental device, although normal reflections may also give rise to (normal) bound states resonating within the QD, these would consist of single-particle states, which do not contribute to the subgap features and, therefore, do not contribute to the transport at small bias voltages. In particular, for the geometry we have considered, the level-spacing of those bound states is $\Delta E = \hbar v\pi/L \approx 9$ meV, which is much larger than the SC gap.

---

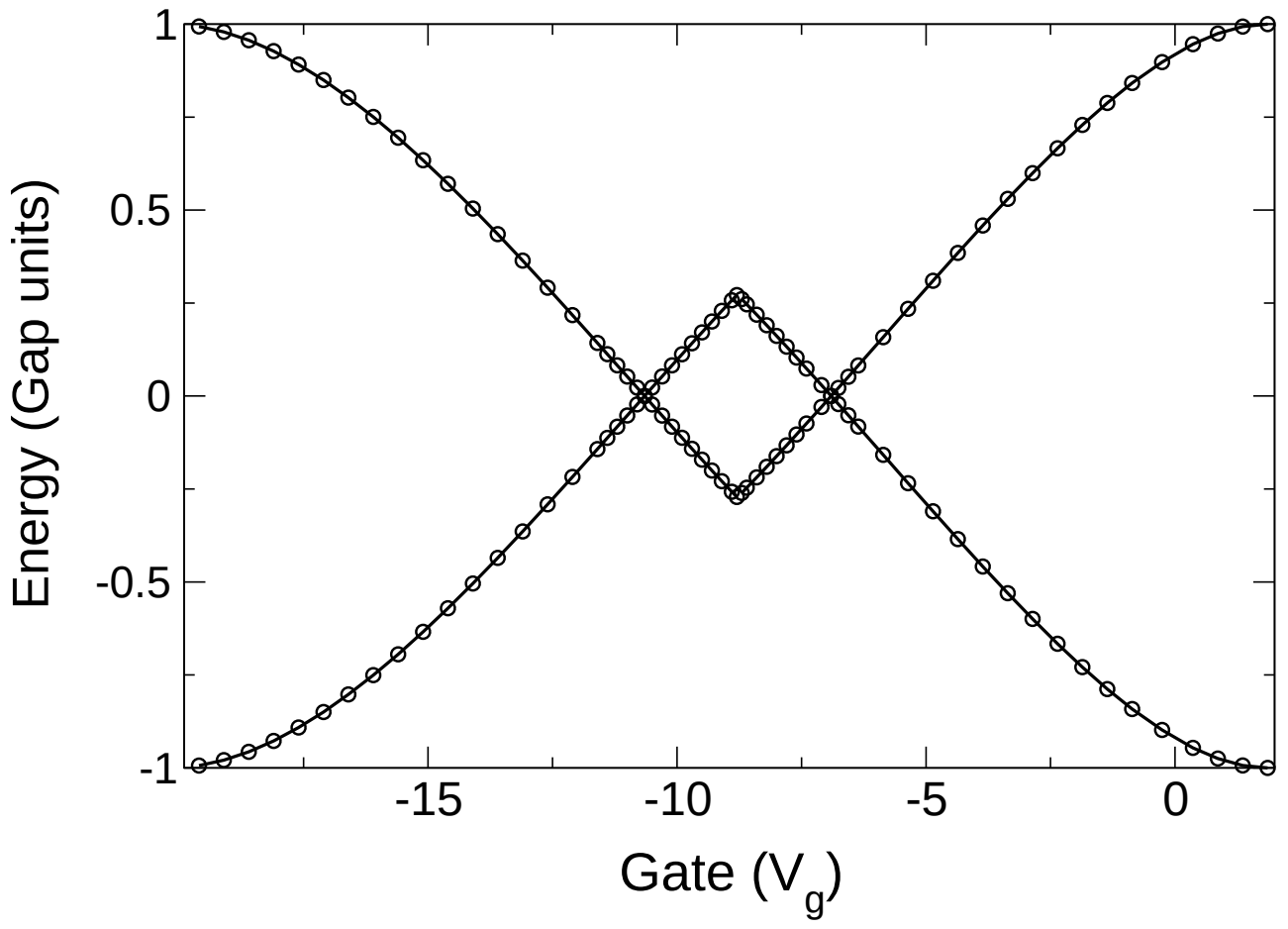

Figure 4: Energy dependence of the lowest energy Andreev bound state within the SC gap with the applied gate voltage. The energy is normalized by the gap, $|\Delta|$, and gate voltage is in V. Barrier height: $V = 0.5$ eV; well depth: $u_0 = 136$ meV; SC gap: $|\Delta| = 1.5$ meV, and charging energy $U = 3.2\,|\Delta|$.